\documentclass[preprint,12pt]{elsarticle}

\usepackage{amssymb, amsmath}
\usepackage{graphicx}
\usepackage{hyperref}
\usepackage{listings}

\journal{Computer Physics Communications}

\begin{document}

\begin{frontmatter}

\title{GWPTools: An extensible Wolfram Language framework for time-dependent Gaussian wave packet dynamics}

\author[1]{Jeremy B. Maddox\corref{cor1}}
\ead{jeremy.maddox@wku.edu}
\cortext[cor1]{Corresponding author}

\author[2]{Ismail Gilani}
\author[2]{Carter M. Heck}
\author[2]{Connor S. Mattingly}

\address[1]{Department of Chemistry and Biochemistry, Western Kentucky University, Bowling Green, KY, 42101-1079, United States}

\address[2]{The Carol Martin Gatton Academy of Mathematics and Science, Western Kentucky University, Bowling Green, KY, 42101-1031, United States}

\begin{abstract}
Gaussian wave packets provide an exact analytical foundation for modeling time-dependent quantum phenomena, effectively bridging the gap between classical trajectories and full quantum mechanics. We introduce GWPTools, a highly extensible Wolfram Language framework designed to compute exact time-dependent Gaussian wave packet dynamics using a unified symbolic-numeric architecture. Built upon a robust ``parameter bus'' and polymorphic dispatch, the software serves as an interactive pedagogical tool for exploring quantum theory and an exact algebraic engine for rigorously benchmarking numerical grid-based and trajectory methodologies. Version 1.0.0 focuses on exact time evolution in driven one-dimensional quadratic potentials. Native extensions into quantum hydrodynamics and phase space representations enable exact energy decompositions, analytical Bohmian trajectories, and symbolic derivations of quasi-probability distributions and information metrics. We demonstrate the framework's modular scalability by prototyping interference in superpositions, open-system decoherence, and coupled two-dimensional oscillators. Ultimately, these capabilities establish a versatile foundation for maximizing the reach of exact analytical modeling toward increasingly complex driven quantum systems.
\end{abstract}

\begin{keyword}
Gaussian wave packets \sep time-dependent quantum mechanics \sep Wolfram Language \sep symbolic computation \sep quantum hydrodynamics \sep phase space
\end{keyword}

\end{frontmatter}

\noindent
{\bf PROGRAM SUMMARY}
  
\begin{small}
\noindent
{\em Program Title:} GWPTools \\
{\em CPC Library link to program files:} \\
{\em Developer's repository link:} \url{https://github.com/maddoxjb/GWPTools}   \\
{\em Licensing provisions:} MIT   \\
{\em Programming language:} Wolfram Language (Mathematica 14.0+)  \\
{\em Supplementary material:} \texttt{GWPTools\_CPC-examples.nb} and its corresponding PDF comprise a supplementary Mathematica notebook detailing the examples presented in this paper. \texttt{GWPVerification.pdf} provides a static record of the program's automated diagnostic suite. \\
{\em Nature of problem:} \\
Numerical modeling of time-dependent quantum systems requires exact benchmark solutions for validation. Furthermore, extending foundational wave packet dynamics to incorporate advanced phase-space mappings, quantum hydrodynamics, and superposition states often yields fragmented, single-purpose codebases. These purely numerical tools lack the continuous symbolic evaluation capabilities necessary to avoid grid artifacts and truncation errors. Moreover, the technical overhead and fragmented nature of these research-grade codebases create significant barriers for interactive pedagogical exploration by students.\\
{\em Solution method:} \\
\texttt{GWPTools} provides an exact symbolic-numeric engine for time-dependent Gaussian wave packets within a highly extensible \textit{Wolfram Language} framework. Utilizing a decoupled parameter bus architecture, the package evaluates exact analytical dynamics in one-dimensional quadratic potentials. The robust polymorphic API instantly generates continuous quantum observables, phase-space distributions, and analytical hydrodynamic trajectories, bypassing numerical solvers entirely. Built for extensibility, the framework successfully prototypes experimental subpackages---including open-system decoherence, multidimensional coupling, and interference---without requiring modifications to the core architectural logic. \\
{\em Additional comments including Restrictions and unusual features:} \\
To guarantee closed-form, purely symbolic evaluation, the Version 1.0.0 core restricts the external potential fields to exact quadratic forms. However, an experimental (i.e., developmental) numerical extension is included to seamlessly integrate arbitrary time-dependent piecewise functions and pulse sequences via memoized differential equation solvers, bypassing strict analytical constraints. \\
{\em References:} \\
{[1]} D. J. Tannor, Introduction to Quantum Mechanics: A Time-Dependent Perspective, University Science Books, Sausalito, CA, 2007.
\end{small}


\section{Introduction}
\label{sec:intro}
Gaussian wave packets (GWPs) are a cornerstone in theoretical chemical physics for modeling time-dependent quantum phenomena. Rooted in exact solutions to the time-dependent Schr\"odinger equation for locally quadratic potentials \cite{Blinder1968}, the formalism was pioneered by Heller \cite{Heller1975, Heller1981} and later elegantly formalized in Tannor's canonical text \cite{Tannor2007}. This theoretical foundation provides an intuitive analytical lens for studying wave packet dynamics, phase space transport, and quantum interference (i.e., superpositions of GWPs). 

The ongoing vitality of this framework is evidenced by rapid contemporary advancements in GWP formulations and propagation techniques. These encompass single-Hessian variants \cite{Barbiero2026} and Hagedorn wave packet basis sets for modeling resonance Raman spectroscopy \cite{Barbiero2026a}, thawed Gaussian Ehrenfest dynamics for mixed quantum-classical systems \cite{Vanicek2026}, structure-preserving geometric integrators \cite{Fereidani2023}, explicitly correlated multidimensional basis sets \cite{Schrader2025}, and coupled formulations designed to improve simulation convergence \cite{CruzRodriguez2025}, alongside unified nonlinear Schr\"odinger equation frameworks \cite{Vanicek2023}. Underpinning this remarkable versatility is the localized mathematical structure of GWPs; by inherently bridging classical phase-space trajectories with quantum wave dynamics, they remain an essential class of model solutions to the time-dependent Schr\"odinger equation \cite{Heller2006}.

While numerical methods for time-dependent quantum mechanics have seen explosive growth \cite{Meyer1990, Johansson2012, Akimov2016, Mattos}, the software infrastructure for exact analytical approaches remains comparatively less organized. The existing computational ecosystem is overwhelmingly tailored toward ever-larger numerical simulations, which presents two distinct challenges. First, grid-based multidimensional frameworks (e.g., MCTDH \cite{Beck2000, Diemer2026}) and comprehensive dynamics suites \cite{Worth2020, Lambert2026, Shakiba2022} provide unparalleled accuracy, but their highly optimized architectures inherently obscure the underlying physics from a pedagogical and analytical perspective. Second, trajectory-based methodologies derived from alternative formulations of quantum mechanics---such as Bohmian mechanics \cite{Bohm1952a, Bohm1952, Holland1995}, interacting quantum trajectories (IQT) \cite{Poirier2010, Schiff2012, Hackebill2026}, and many interacting worlds (MIW) \cite{Hall2014}---must carefully navigate severe numerical instabilities when applied computationally to bypass the exponential scaling limits of rigid grids. Consequently, the diverse numerical implementations \cite{Wyatt2005} and subsequent stabilization techniques developed for these frameworks---ranging from global energy-conserving approximations \cite{Garashchuk2003} and artificial viscosity regularizations \cite{Pauler2004} to dynamically coupled ensembles \cite{Sturniolo2018, Dupuy2022, Chen2023, Maddox2026}---require robust continuous verification. Ultimately, despite their disparate architectures, practitioners across all of these methodologies share a common bottleneck: the need to rigorously validate their algorithms and uncover the obscured physics using exact ground-truth baselines.

To address this critical gap and replace the reliance on fragmented, homegrown validation scripts, we introduce \texttt{GWPTools}. Built upon a strictly decoupled, modular architecture within the \textit{Wolfram Language} \cite{WolframResearch}, the framework is designed to fulfill a three-fold purpose. First, it serves as an intuitive pedagogical platform; by leveraging robust symbolic and interactive visualization capabilities, it enables students to explore exact time-dependent quantum theory without writing complex numerical solvers. Second, it functions as a rigorous benchmarking engine capable of generating exact datasets and continuous boundary conditions to validate numerical methodologies. Finally, by establishing a scalable algebraic foundation, we envision \texttt{GWPTools} evolving into a rapid proof-of-principle calculator that enables researchers to analytically vet their models, ultimately paving the way for future extensions into open quantum systems \cite{Breuer2002} and time-resolved spectroscopy \cite{Mukamel1995}.

\section{Theoretical Framework}
\label{sec:theory}

\subsection{The Four-Eight-Six Parameterization}
\label{subsec:equations_of_motion}
The algebraic core of \texttt{GWPTools} is built directly upon the exact time-dependent Gaussian ansatz. Following the canonical parameterization detailed by Tannor \cite{Tannor2007}, the mathematical form of a generalized, one-dimensional GWP is constructed as:
\begin{align}
\Psi(x,t) = N \exp\left(-\alpha_t(x-x_t)^2 + \frac{i}{\hbar}p_t(x-x_t) + \frac{i}{\hbar}\gamma_t\right)
\end{align}
where $N$ represents a time-independent normalization constant. In its most general mathematical formulation, the four time-dependent parameters $\alpha_t$, $x_t$, $p_t$, and $\gamma_t$ may be fully complex-valued, corresponding to eight real degrees of freedom for their initial values. However, to physically anchor the wave packet to classical observables, these parameters are algebraically reorganized into a strict system of six real-valued degrees of freedom \cite{Tannor2007}. Under this transformation, the coordinate centers $x_t$ and $p_t$ are enforced as strictly real quantities, directly corresponding to the time-dependent position and momentum expectation values.

To guarantee computational stability and purely symbolic representation within the software architecture, the complex shape parameter $\alpha_t$ and the complex global phase $\gamma_t$ are explicitly separated into their real and imaginary components. This structural decomposition maps directly to the core variable structure of \texttt{GWPTools}. While standard users will rarely initialize wave packets with complex coordinate centers, the architecture automatically reduces any generalized eight-parameter input into this physically grounded, six-parameter real representation before time evolution begins.

Substituting the GWP ansatz into the time-dependent Schr\"odinger equation for a quadratic external potential field $V(x,t) = V_0(t) + V_1(t)x + V_2(t)x^2$ yields a coupled set of ordinary differential equations (ODEs) that govern the exact time evolution \cite{Tannor2007}:
\begin{subequations}
\label{eq:eom}
\begin{align}
\dot{\alpha}_t &= -\frac{2i\hbar}{m}\alpha_t^2 + \frac{i}{\hbar}V_2(t) \label{eq:eom_alpha} \\
\dot{x}_t &= \frac{p_t}{m} \label{eq:eom_x} \\
\dot{p}_t &= -(V_1(t) + 2V_2(t)x_t) \label{eq:eom_p} \\
\dot{\gamma}_t &= \frac{p_t^2}{2m} - (V_0(t) + V_1(t)x_t + V_2(t)x_t^2) - \frac{\hbar^2}{m}\alpha_t \label{eq:eom_gamma}
\end{align}
\end{subequations}

While these equations of motion describe the general quadratic system, \texttt{GWPTools} Version 1.0.0 restricts the potential coefficients $V_i(t)$ to a curated library of exact analytical forms to guarantee closed-form, purely symbolic evaluation. Currently, $V_0(t)$ is universally set to zero. Supported configurations include the free particle ($V_1=0, V_2=0$), standard harmonic oscillators ($V_1=0, V_2=\frac{1}{2}m\omega^2$), uniform acceleration ($V_1=-F, V_2=0$), and driven harmonic oscillators with linear or sinusoidal forcing terms (e.g., $V_1(t) = A\sin(\omega t)$). While the core engine relies on these exact forms, the framework also includes an experimental numerical extension capable of evaluating arbitrary time-dependent piecewise functions, pulse sequences, and time-dependent quadratic terms (e.g., Paul-type traps \cite{Paul1990, Feng1995}) via highly optimized, memoized numerical integration with \texttt{NDSolve}.

\subsection{Recursive Polynomial Generation for Exact Observables}
\label{subsec:recursive_generation}
A core computational advantage of \texttt{GWPTools} is its ability to generate exact, arbitrary-order spatial and momentum derivatives without invoking generic, computationally heavy symbolic calculus algorithms. Generic symbolic differentiation of complex exponentials often suffers from severe expression swell and intractable memory costs. Instead, by defining the Gaussian exponent as a quadratic function $Z(x,t)$, any $n$-th spatial derivative can be factored perfectly into a polynomial $P_n(x,t)$ multiplied by the base wavefunction:
\begin{align}
\frac{\partial^n \Psi(x,t)}{\partial x^n} = P_n(x,t)\Psi(x,t)
\end{align}

Because the first spatial derivative $Z'(x,t)$ is strictly linear, the framework constructs these polynomials---which act as generalized Hermite polynomials---using a highly efficient recurrence relation seeded by the trivial base case $P_0(x,t) = 1$. This recursive approach mirrors the highly optimized algorithmic strategies foundational to modern Gaussian basis set evaluations \cite{Obara1986, Helgaker2013}:
\begin{align}
P_n(x,t) = Z'(x,t)P_{n-1}(x,t) + P_{n-1}'(x,t)
\end{align}
From the initial seed, the first derivative naturally yields $P_1(x,t) = Z'(x,t)$, and higher-order terms cascade purely algebraically. This identical recursive generator, when combined with analytical Gaussian moment rules, exactly evaluates arbitrary-order position and momentum expectation values ($\langle x^n \rangle$, $\langle p^n \rangle$) and non-commuting cross-products, entirely bypassing the discrete summation or numerical quadrature typically required in grid-based methodologies \cite{Light2000}.

\subsection{Theoretical Extensions: Quantum Hydrodynamics and Phase Space}
\label{subsec:extensions}
The extensibility of the \texttt{GWPTools} framework is demonstrated by mapping the one-dimensional Gaussian ansatz into complementary physical paradigms, such as quantum hydrodynamics \cite{Takabayasi1952} and phase space representations \cite{Takabayasi1954}. 

The hydrodynamic representation is formulated via Madelung's polar decomposition of the wavefunction, yielding exact continuous fields such as the probability density, velocity fields, and the Fisher information density \cite{Frieden1989, Frieden1998}. Furthermore, \texttt{GWPTools} constructs a perfectly symmetric set of momentum-space hydrodynamic properties \cite{Schuch2023}. This symmetric architecture allows local energy densities to be partitioned exactly into convective (classical) and internal (quantum) components. Central to the internal energy is the quantum potential. Because evaluating the quantum potential (and the quantum force) computationally introduces severe numerical instabilities, its exact analytical evaluation within \texttt{GWPTools} serves as a rigorous testing environment for advanced grid-based solvers \cite{Kendrick2003}, linearized approximations \cite{Garashchuk2003, Garashchuk2004}, Lorentzian basis expansions \cite{Akimov2025}, and complex-plane continuations \cite{Chou2010}.

Additionally, analytical Bohmian trajectory fields are evaluated without numerical integration by mapping physical coordinates to a cumulative probability domain ($c$-space) \cite{Poirier2010, Schiff2012}. This exact algebraic mapping provides the underlying foundation for various wavefunction-free numerical trajectory ensembles \cite{Dupuy2022, Maddox2026}. The spatial trajectory for a specific quantile $c \in (0,1)$ evaluates exactly via the inverse error function:
\begin{equation}
x(c,t) = x_t + \frac{1}{\sqrt{2 \operatorname{Re}(\alpha_t)}} \operatorname{erf}^{-1}(2c-1)
\end{equation}
This continuous formulation provides immediate analytical access to foundational time observables \cite{Lombardini2024}, while the corresponding momentum-space trajectories evaluate with equal analytical efficiency \cite{Schuch2023}.

In addition to hydrodynamics, the architecture transforms states into continuous phase-space distributions \cite{Carruthers1983, Lee1995}. Alongside the exact Wigner function, the software evaluates strictly positive quasi-probability distributions, such as the Husimi representation $Q_{\text{H}}(x,p,t; s)$:
\begin{equation}
Q_{\text{H}}(x,p,t;s) = \iint W(x',p',t) W_{\text{ref}}(x-x', p-p'; s) \,dx'\,dp'
\end{equation}
This enables the exact evaluation of time-dependent phase-space information metrics, such as the continuous Wehrl entropy \cite{Wehrl1978, Lieb1978}. Bypassing numerical quadrature entirely, the exact continuous entropy for a one-dimensional GWP algebraically reduces to a compact, closed-form expression:
\begin{equation}
S_{\text{Wehrl}}(t;s) = 1 + \ln\left( \frac{|\alpha_t + s|^2}{4s \operatorname{Re}(\alpha_t)} \right)
\end{equation}
By providing direct access to this exact theoretical limit, \texttt{GWPTools} serves as a rigorous benchmarking utility for validating discrete numerical phase-space algorithms.

\subsection{Analytical Scope and Planned Extensions}
\label{subsec:analytical_scope}

Finally, the formulations detailed above represent only a foundational subset of the framework's mathematical scope. While the exact single-state 1D parameterization defines the core capabilities of \texttt{GWPTools} Version 1.0.0, the underlying architecture natively accommodates broader physical paradigms. Experimental extensions for multi-component superpositions, multidimensional states, and non-unitary environmental decoherence (via the Caldeira-Leggett master equation \cite{Caldeira1983})---alongside an ensemble module utilizing numerical root-finding for complex trajectory swarms---are currently in active development. Furthermore, an advanced numerical extension is planned to propagate arbitrary time-dependent potential fields; crucially, this module will be uniquely designed to support arbitrary-precision arithmetic, bypassing the standard limitations of machine-precision ODE solvers. To maintain focus on the current release, the formulations and prototypes for these future capabilities are deferred to Section \ref{sec:future_directions} and the Supplementary Information. Conversely, the exhaustive catalog of exact analytical expressions for all Version 1.0.0 features (spanning dynamic observables, energy partitions, and phase-space entropies) is fully integrated and dynamically typeset directly within the software's internal \textit{Wolfram Language} documentation.

\section{Software Architecture}
\label{sec:architecture}

\subsection{The Modular Philosophy and Core Components}
\label{subsec:modular_philosophy}
The architecture of \texttt{GWPTools} is engineered with a strictly decoupled, modular framework that isolates the user-facing frontend dispatcher from the backend mathematical evaluators. This separation of concerns ensures that heavy symbolic algebra and ODE time integration execute independently, without the overhead of continuous user interface validation. Simultaneously, this decoupled design allows developers to seamlessly integrate new physical paradigms directly into the paclet without rewriting the core routing logic. The framework achieves this through six distinct architectural components, detailed in Table \ref{tab:architecture}.

\begin{table}[hbt!]
\centering
\caption{Core architectural components of the \texttt{GWPTools} framework.}
\label{tab:architecture}
\renewcommand{\arraystretch}{1.3} 
\begin{tabular}{l p{0.65\textwidth}}
\hline
\textbf{Component} & \textbf{Description} \\
\hline
\texttt{GWPTools.wl} & The master paclet loader that orchestrates the initialization sequence for the registries, dispatchers, and math engines. \\
\texttt{GWPRegistry.wl} & The central database managing dynamic physical properties and analytical potential models. \\
\texttt{GWPDispatcher.wl} & The polymorphic routing engine and frontend argument-validation manager. \\
\texttt{GWPDeveloper.wl} & The namespace bridge housing the universal parameter dictionary and developer utilities. \\
\texttt{GWPEngineTYPE.wl} & The isolated computational backends that contain the raw algebraic evaluators and parameter buses for specific paradigms (e.g., \texttt{GWPEngine1D.wl}). \\
\texttt{GWPEXT.wl} & High-level orchestrator subpackages that dynamically inject complex, higher-order property classes when their required parent math engine is active (e.g., \texttt{GWPHydrodynamics.wl}). \\
\hline
\end{tabular}
\end{table}

\subsection{Frontend API: The GWP Constructor and Introspection}
\label{subsec:frontend_api}
All standard user (i.e., non-developer) interactions are channeled through exactly two exported symbols: the \texttt{GWP} constructor and the \texttt{GWPObject} data structure. Acting as the primary wave packet constructor, \texttt{GWP} utilizes a curried signature, \texttt{GWP[type][parameters, options]}, where the \texttt{type} string prompts the dispatcher to select the appropriate backend mathematical engine. In addition to initializing states, \texttt{GWP} serves a dual purpose as the primary interface for the framework's dynamic introspection capabilities. By querying the constructor with specific strings---such as \texttt{"Types"}, \texttt{"Classes"}, or \texttt{"Properties"}---users can instantly reveal which wave packet engines are currently loaded in the kernel, the available physical property categories, and their registered macros.

Upon evaluating the input arguments, the constructor parses the parameters and options to generate a \texttt{GWPObject}, a highly structured data container representing the physical state. This design ensures the frontend API remains minimalist, delegating all complex structural routing to the encapsulated \texttt{GWPDispatcher}. When a user requests a property directly from a \texttt{GWPObject}, the dispatcher intercepts the query, retrieves the corresponding metadata from the central registry, and executes the exact backend mathematical signature required for that specific object's state.

\subsection{The Central Registry and Polymorphic Dispatcher}
\label{subsec:registry_dispatch}
The interoperability of the six architectural pillars is governed by a centralized database and a multi-tier polymorphic router. At initialization, \texttt{GWPRegistry.wl} establishes an empty property and potential model database. As specific mathematical engines (e.g., the core 1D analytical engine) or extension subpackages (e.g., hydrodynamics) are loaded into the kernel, they inject their unique capabilities into this registry.

Crucially, when an engine registers a physical observable, it records a strict mathematical signature taxonomy rather than just a property name. Every property is explicitly mapped to both a \texttt{DynamicClass}, which is executed when an external potential model is active, and a \texttt{StaticClass}, for when the potential is not specified. For example, observables leveraging the polynomial generators detailed in Section \ref{subsec:recursive_generation} map their dynamic evaluation to a \texttt{Recursive} signature (requiring the integer order alongside spatial and temporal variables) and their static evaluation to a \texttt{RecursiveSpatial} signature (requiring only the order and spatial coordinates). When a user requests a property, the \texttt{GWPDispatcher} queries this dual-signature metadata and evaluates the object's structural state to seamlessly route the request. This polymorphic dispatch allows the frontend to adapt to complex physical states without hardcoded API overloads, ensuring new engines can be integrated without modifying the underlying routing architecture.

\subsection{The Parameter Bus and Developer Namespace Bridge}
\label{subsec:parameter_bus}
To guarantee seamless interoperability across diverse physical models, every computational engine is built upon a macro-expansion strategy termed the ``Parameter Bus.'' Each GWP type and its corresponding math engine introduce a unique sequence of variables specifically tailored to its physical paradigm. This bus dynamically expands into a strict variable sequence that stores all foundational attributes, physical constants, potential coefficients, and time-origin metadata required to fully define the state.

For the primary 1D analytical engine, the theoretical formulation of the time-dependent wave packet defines the state using complex-valued parameters: the shape parameter $\alpha_t$, the position center $x_t$, the momentum center $p_t$, and the spatially independent phase $\gamma_t$. To optimize computational performance and safely handle branch cuts during exact integration, the \texttt{GWPTools} codebase explicitly separates these into foundational real and imaginary components. The 1D engine utilizes an 11-element sequence where these variables are mapped to internal code symbols: \texttt{RA} and \texttt{IA} (for the real and imaginary components of $\alpha_t$), \texttt{RX} (for $x_t$), \texttt{RP} (for $p_t$), and \texttt{RG} and \texttt{IG} (for the global phase $\gamma_t$), with the remaining five slots reserved for the reduced Planck constant, the system mass, potential coefficients, and time-origin metadata.

A primary challenge in modular computer algebra development is preventing symbol shadowing across different internal package contexts while rigorously avoiding the pollution of the user's global namespace. To resolve this, \texttt{GWPDeveloper.wl} functions as an internal master namespace bridge, explicitly hoisting these foundational symbols into an invisible developer context. By maintaining this centralized dictionary, extension subpackages can act as high-level orchestrators; they natively read the active Parameter Bus arrays supplied by any engine to generate complex physical observables without requiring namespace prefixing.

Furthermore, the backend functions for a given type adhere to a rigid macro-expansion format for utilizing and passing the parameter bus. To manage this data exchange, each engine defines a complementary pair of macros. For example, the 1D engine utilizes \texttt{GWP1DARG} to define the function's argument sequence, paired with \texttt{GWP1DVAL} to subsequently pass the evaluated parameters downstream. This strict format guarantees that parameters are passed reliably between nested backend mathematical functions without breaking delayed evaluation. By strictly encapsulating these shared symbols and internally managing load dependencies via \texttt{Needs}, the framework guarantees a completely silent, collision-free initialization sequence that respects the standard user's global workspace. Should a developer need to interact directly with these backend evaluators for debugging or prototyping new extensions, they can explicitly load the developer context using \texttt{Get} (rather than \texttt{Needs}) to force the hidden mathematical heads onto the active \texttt{\$ContextPath}.

\subsection{Runtime Telemetry and Automated Verification}
\label{subsec:verification}

To assist developers in isolating dependency resolution errors and tracing the package initialization sequence, the framework includes a native runtime telemetry feature. By setting the global variable \texttt{\$GWPDebug = True} prior to loading the paclet, the master orchestrator and its constituent subpackages output verbose console logs during the loading sequence. While currently utilized to diagnose context path isolation and module loading order, this telemetry infrastructure is designed to be expanded in future releases to trace deep algorithmic evaluation pathways during complex multi-step calculations.

To guarantee cross-platform reproducibility and rigorous mathematical accuracy, \texttt{GWPTools} includes a dedicated testing and verification subpackage, \texttt{GWPDiagnostics}. Rather than relying on ad hoc checks, the framework evaluates complex symbolic logic and integration entirely from scratch via an exhaustive suite of internal unit tests. This automated verification process is fully documented and accessible to users through the \texttt{GWPVerification.nb} notebook, located in the package's \texttt{Tests} directory.

The diagnostic suite systematically stress-tests every layer of the architecture. At the frontend, it validates that the API correctly parses overloaded inputs, applies strict domain constraints (such as Heisenberg uncertainty bounds), and accurately routes data to construct the foundational parameter bus. At the backend, it confirms the analytical integrity of the mathematical engines by cross-verifying explicitly coded closed-form derivatives against generic symbolic differentiation up to the sixth order, ensuring strict probability conservation and dimensional consistency across position, momentum, and energy observables. Furthermore, the suite proves the exactness of complex observables by comparing explicitly coded expectation values against full-space symbolic integration. For the advanced hydrodynamic extensions, the diagnostics validate the exact analytical inversion of cumulative distribution functions, ensuring that all quantum potentials, flow velocities, and Bohmian trajectory fields map consistently across coordinate representations.

By consolidating these validations into a single executable notebook that generates standardized \texttt{TestReportObject} datasets, \texttt{GWPTools} provides reviewers and developers with a transparent, one-click guarantee of mathematical exactness. This rigorous foundation ensures that any future extensions to the core physics engines maintain absolute backward compatibility.

\subsection{Integrated Documentation and User Guides}
\label{subsec:documentation}
A primary barrier to the adoption of specialized computational research tools is the absence of accessible, interactive documentation. To address this, \texttt{GWPTools} includes a comprehensive documentation suite natively integrated into the standard \textit{Wolfram Language} Documentation Center. This built-in infrastructure is carefully structured to support users across all levels of expertise. Core frontend symbols, such as \texttt{GWP} and \texttt{GWPObject}, feature interactive, executable reference pages detailing exact syntax and property introspection. For broader workflows, the package provides a series of user-focused Technical Notes that guide researchers through specific physical applications, spanning core mechanics (\texttt{GWPUser}, \texttt{GWPEngine1D}) to advanced analytical frameworks (\texttt{GWPPhaseSpace}, \texttt{GWPHydrodynamics}). Finally, to facilitate community contributions, a dedicated \texttt{GWPDeveloper} Technical Note explicitly outlines the internal Parameter Bus protocols, empowering researchers to safely integrate custom physical models into the backend architecture.

\section{Illustrative Examples}
\label{sec:examples}

Guided by the architectural principles of polymorphic dispatch, symbolic-numeric hybridization, and massive extensibility, \texttt{GWPTools} presents a minimalist and intuitive frontend to the user. The following examples utilize the \textit{Wolfram Language} syntax to demonstrate the practical application of these principles for the Version 1.0.0 release.

\subsection{Framework Introspection and Central Registry}
\label{subsec:ex_introspection}
The \texttt{GWP} constructor serves as the primary gateway for discovering the framework's capabilities. A user can dynamically query the active kernel to reveal available engine types, property classes, and specific macros without needing to reference external documentation:
\begin{lstlisting}[language=Mathematica, basicstyle=\footnotesize\ttfamily, columns=fullflexible, keepspaces=true, breaklines=true]
In[1]:= Needs["GWPTools`"]
In[2]:= GWP["Types"]

Out[2]= {"1D"}

In[3]:= GWP["1D"]["Classes"]

Out[3]= {"DynamicParameters", "Energies", "ExpectationValues", 
         "Probabilities", "StaticParameters", "Wavefunctions"}
\end{lstlisting}
Filtering by a specific class yields the exact macros required for evaluation. For instance, querying the \texttt{"Probabilities"} class reveals the available density and cumulative distribution functions:
\begin{lstlisting}[language=Mathematica, basicstyle=\footnotesize\ttfamily, columns=fullflexible, keepspaces=true, breaklines=true]
In[4]:= GWP["1D"]["Properties", "Probabilities"]

Out[4]= {"CumulativeDistributionE", "CumulativeDistributionP", 
         "CumulativeDistributionX", "DensityE", "DensityP", 
         "DensityX", "ProbabilityE", "ProbabilityP", "ProbabilityX"}
\end{lstlisting}
To understand how the frontend routes property requests to the backend evaluators, a user can query the raw registry records for either a broad property class or an individual property. For example, querying the records for \texttt{"DensityX"} returns a structured association containing the exact signature metadata for that property:
\begin{lstlisting}[language=Mathematica, basicstyle=\footnotesize\ttfamily, columns=fullflexible, keepspaces=true, breaklines=true]
In[5]:= GWP["1D"]["Records", "DensityX"]

Out[5]= {<|"Property" -> "DensityX", "ShortKey" -> "RHOX", 
  "Class" -> "Probabilities", "DynamicClass" -> "Recursive", 
  "StaticClass" -> "RecursiveSpatial"|>}
\end{lstlisting}
Table \ref{tab:records_dataset} details the corresponding signature metadata for the entire \texttt{"Probabilities"} property class, demonstrating how each observable maps to a \texttt{"ShortKey"} abbreviation alongside its specific \texttt{"DynamicClass"} and \texttt{"StaticClass"} routing strings. 

Similarly, users can introspect the physical environments available to the engine. Querying the active potentials with the string \texttt{"PotentialNames"} returns a list of all currently registered analytical potential models:
\begin{lstlisting}[language=Mathematica, basicstyle=\footnotesize\ttfamily, columns=fullflexible, keepspaces=true, breaklines=true]
In[6]:= GWP["1D"]["PotentialNames"]

Out[6]= {"Free", "Harmonic", "HO", "Linear", "LinearForcedHO", 
  "NonResonantForcedHO", "ParabolicBarrier", "ResonantForcedHO"}
\end{lstlisting}
A subsequent \texttt{"PotentialRecords"} query can then retrieve the metadata governing these models, including the expected input templates for parameterization and the internal backend symbols responsible for their evaluation.

\begin{table}[hbt!]
\centering
\caption{Tabular representation of the registry records returned for the \texttt{"Probabilities"} class. The \texttt{GWPDispatcher} utilizes the \texttt{DynamicClass} and \texttt{StaticClass} keys to automatically route user queries to the correct mathematical signature based on the presence of an external potential.}
\label{tab:records_dataset}
\renewcommand{\arraystretch}{1.3}
\begin{tabular}{lllll}
\hline
\textbf{Property} & \textbf{ShortKey} & \textbf{Class} & \textbf{DynamicClass} & \textbf{StaticClass} \\
\hline
DensityX & RHOX & Probabilities & Recursive & RecursiveSpatial \\
DensityP & RHOP & Probabilities & Recursive & RecursiveSpatial \\
DensityE & RHOE & Probabilities & Field & Spatial \\
CumulativeDistributionX & CX & Probabilities & Field & Spatial \\
CumulativeDistributionP & CP & Probabilities & Field & Spatial \\
CumulativeDistributionE & CE & Probabilities & Field & Spatial \\
ProbabilityX & PROBX & Probabilities & Bivariate & BivariateSpatial \\
ProbabilityP & PROBP & Probabilities & Bivariate & BivariateSpatial \\
ProbabilityE & PROBE & Probabilities & Bivariate & BivariateSpatial \\
\hline
\end{tabular}
\end{table}

\subsection{Symbolic-Numeric Hybridization and Dynamic Signatures}
\label{subsec:ex_hybrid_dispatch}
\begin{figure}[htbp]
    \centering
    \includegraphics[width=0.75\textwidth]{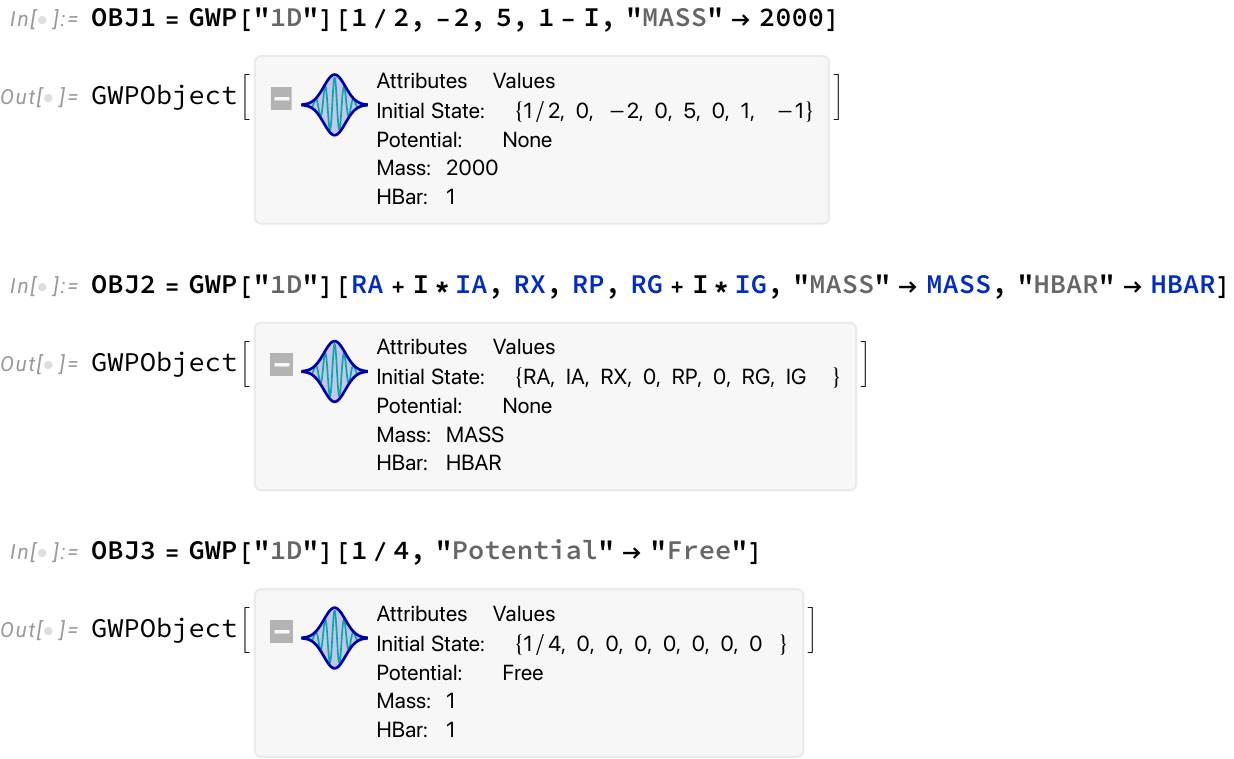}
    \caption{Visual representation of the \texttt{GWPObject} summary boxes generated by the \texttt{GWP} constructor for varying input parameters. The frontend interface provides immediate access to the top-level state metadata, while the underlying data container stores the full 11-element parameter bus utilized by the polymorphic dispatcher.}
    \label{fig:gwp_box}
\end{figure}

State initialization is handled identically regardless of whether the user inputs exact rational numbers, machine-precision floats, or purely symbolic variables. By default, the \texttt{GWP} constructor returns a visually formatted summary box containing the state's top-level metadata (see Figure \ref{fig:gwp_box} for the \texttt{OBJ1}, \texttt{OBJ2}, and \texttt{OBJ3} instances). However, the underlying \texttt{GWPObject} data container explicitly stores the expanded Parameter Bus utilized for downstream routing. 

Recalling the theoretical parameterization from Section \ref{subsec:parameter_bus}, the constructor accepts a sequence of optional positional arguments---shape, position, momentum, and phase---followed by an option pattern. Unspecified parameters automatically fall back to default values ($\alpha_0=1/4$ and zero for $x_0$, $p_0$, and $\gamma_0$), while system constants default to natural units (e.g., $\hbar=1$) and the potential defaults to \texttt{None}. We demonstrate this flexibility by fully specifying an exact numeric state with a custom mass (\texttt{OBJ1}), generating a completely symbolic state (\texttt{OBJ2}), and initializing a time-dependent free particle (\texttt{OBJ3}, which provides only the shape parameter and relies on defaults for the remaining sequence):
\begin{lstlisting}[language=Mathematica, basicstyle=\footnotesize\ttfamily, columns=fullflexible, keepspaces=true, breaklines=true]
In[7]:= OBJ1 = GWP["1D"][1/2, -2, 5, 1 - I, "MASS" -> 2000];
In[8]:= OBJ2 = GWP["1D"][RA + I*IA, RX, RP, RG + I*IG, 
                         "MASS" -> MASS, "HBAR" -> HBAR];
In[9]:= OBJ3 = GWP["1D"][1/4, "Potential" -> "Free"];
\end{lstlisting}

Evaluating the spatial probability density (\texttt{"DensityX"}) seamlessly adapts to the input type. The exact numerical object yields an exact reduced fraction, while the symbolic object returns the analytical formula:
\begin{lstlisting}[language=Mathematica, basicstyle=\footnotesize\ttfamily, columns=fullflexible, keepspaces=true, breaklines=true]
In[10]:= OBJ1["DensityX"][x]

Out[10]= E^-(2 + x)^2/Sqrt[\[Pi]]

In[11]:= rhox = OBJ2["DensityX"][x]

Out[11]= E^(-2 RA (-RX + x)^2) Sqrt[2/\[Pi]] Sqrt[RA]
\end{lstlisting}

Because \texttt{GWPTools} isolates the physical parameters, it can automatically generate a rigorous set of logical domain constraints. Querying the \texttt{"Assumptions"} property retrieves these constraints, which enforce real-valued boundaries and physical limits (e.g., strictly positive mass). Feeding these automatically generated assumptions into the native \textit{Wolfram Language} \texttt{Integrate} function eliminates zero-variance singularities and branch-cut checks, dramatically accelerating symbolic evaluations:
\begin{lstlisting}[language=Mathematica, basicstyle=\footnotesize\ttfamily, columns=fullflexible, keepspaces=true, breaklines=true]
In[12]:= assum = OBJ2["Assumptions"]

Out[12]= (HBAR | IA | IG | MASS | RA | RG | RP | RX) \[Element] Reals 
         && RA > 0 && HBAR > 0 && MASS > 0

In[13]:= Timing[Integrate[rhox, {x, -\[Infinity], \[Infinity]}]]

Out[13]= {7.79101, ConditionalExpression[1, Re[RA] >= 0]}

In[14]:= Timing[Integrate[rhox, {x, -\[Infinity], \[Infinity]}, 
                         Assumptions -> assum]]

Out[14]= {0.373837, 1}
\end{lstlisting}

The power of the polymorphic dispatcher is further evident when an active potential is introduced. For \texttt{OBJ3}, the dispatcher automatically routes the \texttt{"DensityX"} query through the \texttt{DynamicClass} signature, returning an exact function of both space and time ($x, t$) rather than just strictly space ($x$):
\begin{lstlisting}[language=Mathematica, basicstyle=\footnotesize\ttfamily, columns=fullflexible, keepspaces=true, breaklines=true]
In[15]:= OBJ3["DensityX"][x, t] // Simplify

Out[15]= E^(-((2 x^2)/(4 + t^2))) Sqrt[2/\[Pi]] Sqrt[1/(4 + t^2)]
\end{lstlisting}

\subsection{Grid Mapping for Numerical Benchmarking}
\label{subsec:ex_benchmarking}
A primary application of the exact algebraic results generated by \texttt{GWPTools} is the rigorous benchmarking of numerical grid-based methods or trajectory solvers. Rather than producing discrete floating-point approximations, the framework outputs exact continuous functions. Furthermore, because the frontend API natively supports \textit{Wolfram Language} listability, users can map these exact properties across discretized arrays. This combination provides an absolute, uncorrupted ground truth for calculating numerical truncation errors.

For instance, mapping the position uncertainty function of the freely dispersing wave packet (\texttt{OBJ3}) over a discrete time array yields an exact list of surds, which can subsequently be evaluated to arbitrary precision:
\begin{lstlisting}[language=Mathematica, basicstyle=\footnotesize\ttfamily, columns=fullflexible, keepspaces=true, breaklines=true]
In[16]:= OBJ3["PositionUncertainty"]@Range[0, 10, 2]

Out[16]= {1, Sqrt[2], Sqrt[5], Sqrt[10], Sqrt[17], Sqrt[26]}

In[17]:= N[\%, 18]

Out[17]= {1.00000000000000000, 1.41421356237309505, 
          2.23606797749978970, 3.16227766016837933, 
          4.12310562561766055, 5.09901951359278483}
\end{lstlisting}

For broader time-dependent simulations, these exact analytical functions can be evaluated across dense spatio-temporal grids. Utilizing native functions like \texttt{Outer}, users can rapidly generate exact trajectory matrices without accumulating the integration errors inherent to standard differential equation solvers:
\begin{lstlisting}[language=Mathematica, basicstyle=\footnotesize\ttfamily, columns=fullflexible, keepspaces=true, breaklines=true]
In[18]:= tdat = Range[0, 5, 1/5];
         xdat = Range[-5, 5, 1/4];
         data = Outer[OBJ3["DensityX"][#1, #2] &, xdat, tdat];
         Dimensions[data]

Out[18]= {41, 26}
\end{lstlisting}
Because the generated \texttt{data} matrix retains its exact symbolic formulation, individual grid points can be isolated and evaluated perfectly without standard precision limits:
\begin{lstlisting}[language=Mathematica, basicstyle=\footnotesize\ttfamily, columns=fullflexible, keepspaces=true, breaklines=true]
In[19]:= data[[21, 1]]

Out[19]= 1/Sqrt[2 \[Pi]]

In[20]:= N[data[[21, 1]], 20]

Out[20]= 0.39894228040143267794
\end{lstlisting}

\begin{figure}[htbp]
    \centering
    \includegraphics[width=0.6\textwidth]{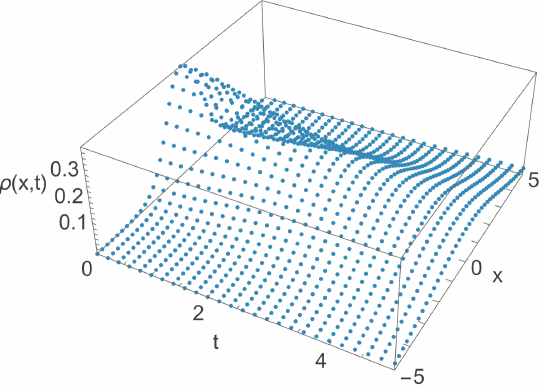}
    \caption{A three-dimensional visualization of the exact spatio-temporal probability density grid (\texttt{data}) generated for the freely dispersing Gaussian wave packet (\texttt{OBJ3}). Because the data is generated algebraically, the surface avoids the numerical accumulation errors typical of discretized time-evolution solvers.}
    \label{fig:grid_mapping}
\end{figure}

Beyond generating standard visualizations---such as the \texttt{ListPointPlot3D} free-particle dispersion surface shown in Figure \ref{fig:grid_mapping}---the primary utility of these exact algebraic matrices is rigorous algorithmic validation. Because the generated arrays represent a zero-error reference standard, developers can directly utilize them for benchmark testing by computing point-by-point residuals (or exact errors) against their own numerically generated datasets. This allows for precise error analysis, enabling developers to establish rigorous stability tolerances and verify the convergence rates of new grid-based or trajectory PDE solvers.

\subsection{Hydrodynamics and Phase Space Extensions}
\label{subsec:ex_extensibility}
\begin{figure}[htbp]
    \centering
    \begin{minipage}{0.48\textwidth}
        \centering
        \includegraphics[width=\linewidth]{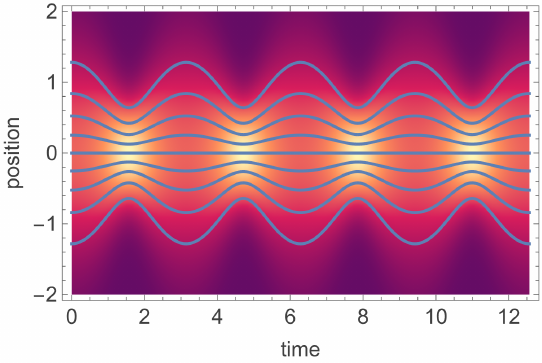}
        \centerline{(a)}
    \end{minipage}\hfill
    \begin{minipage}{0.48\textwidth}
        \centering
        \includegraphics[width=\linewidth]{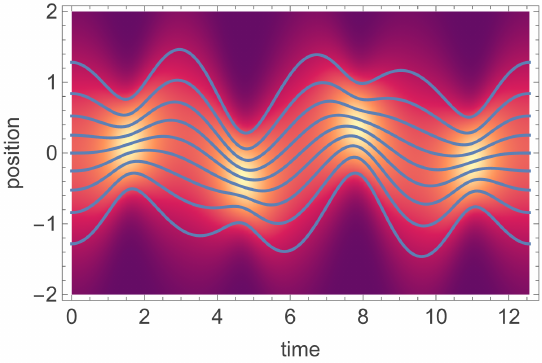}
        \centerline{(b)}
    \end{minipage}
    \caption{Exact Bohmian trajectories overlaid on the time-dependent spatial probability density $\rho(x,t)$. (a) Evolution of a Gaussian wave packet within a standard harmonic oscillator potential (\texttt{"HO"}), demonstrating periodic breathing and stable kinematics. (b) Complex driven dynamics within a non-resonant forced harmonic oscillator (\texttt{"NonResonantForcedHO"}). In both cases, the hydrodynamic trajectories---corresponding to probability quantiles $c \in \{1/10, 2/10, \dots, 9/10\}$---are generated algebraically by directly evaluating the exact inverse cumulative distribution function (\texttt{"XC"}). This analytical mapping completely bypasses the need for numerical differential equation solvers, guaranteeing exact, infinite-precision trajectory fields for any specified time domain.}
    \label{fig:hydro_trajectories}
\end{figure}

The framework is engineered to horizontally scale by loading specialized subpackages, allowing users to extract advanced physical observables with a single query. To demonstrate this dynamic extensibility, loading the \texttt{GWPHydrodynamics} and \texttt{GWPPhaseSpace} modules instantly injects new property classes directly into the central registry:
\begin{lstlisting}[language=Mathematica, basicstyle=\footnotesize\ttfamily, columns=fullflexible, keepspaces=true, breaklines=true]
In[21]:= Needs["GWPTools`GWPHydrodynamics`"]
         Needs["GWPTools`GWPPhaseSpace`"]

In[22]:= GWP["1D"]["Classes"]

Out[22]= {"BohmianTrajectories", "DynamicParameters", "Energies", 
          "ExpectationValues", "HydrodynamicsP", "HydrodynamicsX", 
          "PhaseSpace", "Probabilities", "StaticParameters", 
          "Wavefunctions"}
\end{lstlisting}
Introspecting these newly registered classes reveals a comprehensive suite of advanced theoretical observables that are now instantly available to the frontend polymorphic dispatcher:
\begin{lstlisting}[language=Mathematica, basicstyle=\footnotesize\ttfamily, columns=fullflexible, keepspaces=true, breaklines=true]
In[23]:= GWP["1D"]["Properties", "HydrodynamicsX"]

Out[23]= {"AmplitudeX", "ConvectiveKineticDensityX", "CurrentX", 
          "FisherInformationDensityX", "InternalKineticDensityX", 
          "MomentumFieldX", "OsmoticVelocityX", "PhaseX", "QuantumForceX", 
          "QuantumPotentialX", "QuantumStressX", "TotalEnergyDensityX", 
          "TotalKineticDensityX", "TotalPotentialDensityX", "VelocityX"}

In[24]:= GWP["1D"]["Properties", "PhaseSpace"]

Out[24]= {"DensityMatrixP", "DensityMatrixX", "HusimiQ", "Purity", 
          "VonNeumannEntropy", "WehrlEntropy", "WignerFunction"}
\end{lstlisting}
With the registry expanded, the framework enables the exact algebraic extraction of complex properties that typically require extensive manual derivation. For instance, exact Bohmian trajectories can be evaluated without relying on numerical differential equation solvers. By mapping physical coordinates to a cumulative probability domain ($C$-space), trajectories for specific probability quantiles are retrieved as exact analytical functions. We can initialize a standard harmonic oscillator (\texttt{OBJ1}) alongside a non-resonant forced oscillator (\texttt{OBJ2}) and instantly extract their trajectory fields (\texttt{"XC"}):
\begin{lstlisting}[language=Mathematica, basicstyle=\footnotesize\ttfamily, columns=fullflexible, keepspaces=true, breaklines=true]
In[25]:= cdat = Range[1/10, 9/10, 1/10];
         OBJ1 = GWP["1D"]["Potential" -> "HO"];
         OBJ2 = GWP["1D"]["Potential" -> {"NonResonantForcedHO", 1, 1/5, 3/2}];
         traj1 = OBJ1["XC"][cdat, t];
         traj2 = OBJ2["XC"][cdat, t];
\end{lstlisting}
Because these outputs are continuous analytical functions, they can be trivially passed to native visualization functions to generate high-resolution spatio-temporal datasets and trajectory graphs, as demonstrated in Figure \ref{fig:hydro_trajectories}.

Similarly, the phase space extension grants immediate access to advanced quasi-probability distributions and quantum information metrics. While standard packages limit users to fixed numerical grids, \texttt{GWPTools} analytically derives these representations while carrying arbitrary symbolic coarse-graining parameters. For example, generating the Husimi Q-representation for a static wave packet (\texttt{OBJ4}) parameterized by an arbitrary shape parameter $s$ instantly enables the exact computation of the continuous Wehrl entropy:
\begin{lstlisting}[language=Mathematica, basicstyle=\footnotesize\ttfamily, columns=fullflexible, keepspaces=true, breaklines=true]
In[26]:= OBJ4 = GWP["1D"][];

In[27]:= Simplify[OBJ4["HusimiQ", s][x, p], s > 0]

Out[27]= (2 E^(-((2 (p^2 + s x^2))/(1 + 4 s))) Sqrt[s])/(\[Pi] + 4 \[Pi] s)

In[28]:= OBJ4["WehrlEntropy", s]

Out[28]= 1 + Log[(1/4 + s)^2/s]
\end{lstlisting}
These examples highlight how the framework bridges the gap between simple state initialization and the extraction of complex theoretical observables, providing users with a robust algebraic toolset that easily accommodates arbitrary symbolic parameters.

\section{Framework Scalability and Advanced Prototypes}
\label{sec:future_directions}

The robust, modular API developed for \texttt{GWPTools} was explicitly engineered for two distinct axes of scalability. Horizontal scalability is achieved by dynamically loading specialized extension subpackages (such as the hydrodynamics and phase space modules demonstrated in Section \ref{subsec:ex_extensibility}) to expand the observable capabilities of an existing state. Conversely, vertical scalability is realized through the integration of entirely new mathematical engines that redefine the foundational physical state itself. While the Version 1.0.0 release focuses on single-state 1D dynamics, the architectural scaffolding natively supports vertical expansion into advanced state formulations, paving the way for a comprehensive suite of quantum dynamics and theoretical chemistry applications.

\subsection{Superposition States}
\label{subsec:superposition}
\begin{figure}[htbp]
    \centering
    \begin{minipage}{0.48\textwidth}
        \centering
        \includegraphics[width=\linewidth]{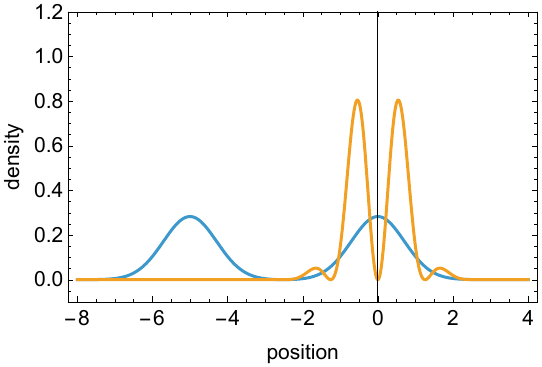}
        \centerline{(a) Antisymmetric ($t=0, \pi/2$)}
    \end{minipage}\hfill
    \begin{minipage}{0.48\textwidth}
        \centering
        \includegraphics[width=\linewidth]{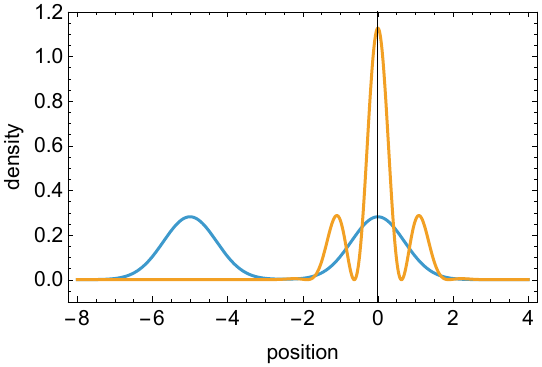}
        \centerline{(b) Symmetric ($t=0, \pi/2$)}
    \end{minipage}
    
    \vspace{0.5cm} 
    
    \begin{minipage}{0.48\textwidth}
        \centering
        \includegraphics[width=\linewidth]{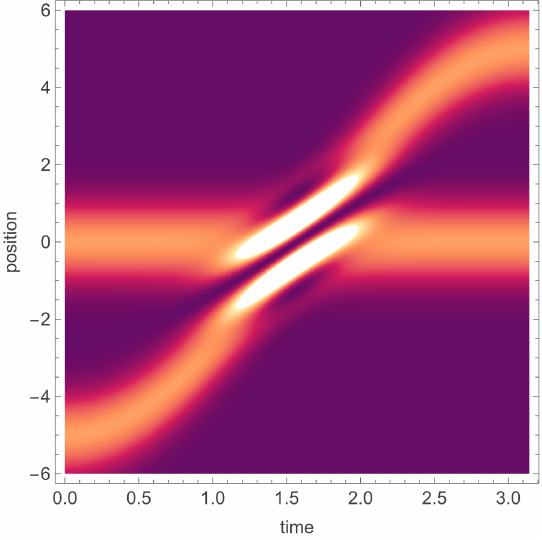}
        \centerline{(c) Antisymmetric Density}
    \end{minipage}\hfill
    \begin{minipage}{0.48\textwidth}
        \centering
        \includegraphics[width=\linewidth]{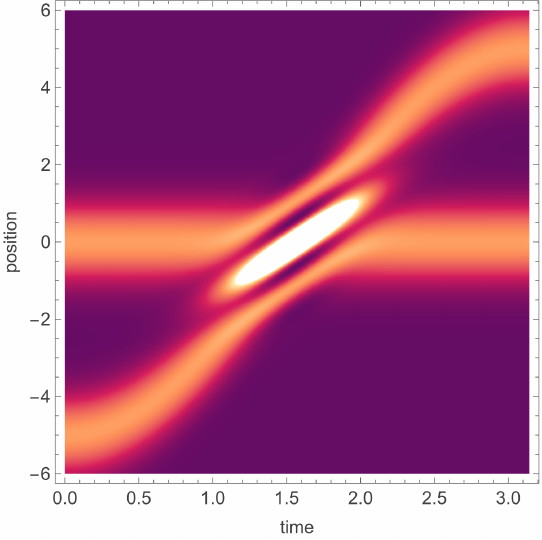}
        \centerline{(d) Symmetric Density}
    \end{minipage}
    
\caption{Exact analytical dynamics of superposition states within a harmonic oscillator potential. The panels illustrate a displaced coherent state wave packet colliding with a stationary ground-state wave packet. Panels (a) and (c) display the antisymmetric superposition, showcasing the persistent probability node at the origin during the collision event ($t=\pi/2$). Panels (b) and (d) display the symmetric superposition, demonstrating constructive interference at the collision center.}
    \label{fig:advanced_dynamics}
\end{figure}

Foremost among these planned extensions is the integration of exact superposition states (\texttt{"SS1D"} type) for studying complex wave packet interference and ultimately entangled particle dynamics. To demonstrate this capability, we configure a scenario where a displaced coherent state collides with a stationary ground-state wave packet within a harmonic oscillator potential. Using the prototype \texttt{GWPEngineSS1D} module, we initialize both antisymmetric and symmetric superposition states, and then extract their exact time-dependent spatial densities:
\begin{lstlisting}[language=Mathematica, basicstyle=\footnotesize\ttfamily, columns=fullflexible, keepspaces=true, breaklines=true]
In[29]:= Needs["GWPTools`GWPEngineSS1D`"]

In[30]:= ARGS = Sequence[{1/2, 1/2}, {-5, 0}, "Potential" -> "HO"];
         OBJ1 = GWP["SS1D"][ARGS, "C" -> -1]; (* anti-symmetric *)
         OBJ2 = GWP["SS1D"][ARGS, "C" -> 1];  (* symmetric *)

In[31]:= rhoss1[x_, t_] = OBJ1["DensityX"][x, t];
         rhoss2[x_, t_] = OBJ2["DensityX"][x, t];
\end{lstlisting}
Evaluating these continuous density functions throughout the collision reveals highly intricate, exact interference patterns. As illustrated in Figure \ref{fig:advanced_dynamics}, this analytical approach effortlessly captures the distinct topological features of the different states. Because the evaluations bypass numerical grid resolution limits, the framework precisely preserves delicate spatial structures that are often smeared by standard grid solvers---such as the perfectly persistent probability node at the origin for the antisymmetric state and the sharp constructive interference peak for the symmetric state.

\subsection{Open Quantum Systems and Dissipation}
\label{subsec:dissipation}
The framework is also being actively extended to model open quantum systems via reduced density matrix calculations (\texttt{"RDM1D"} type). Loading the prototype \texttt{GWPEngineRDM1D} module allows for the exact algebraic propagation of density matrices according to the Caldeira-Leggett master equation \cite{Caldeira1983}. To parameterize the environment, the engine accepts a specialized potential template, \texttt{\{"HarmonicCL", $\omega$, $\gamma$, kT\}}. In the example below, we initialize an underdamped oscillator with $\omega=1$ that is coupled to a thermal bath with thermal energy $kT=1$ and a damping rate of $\gamma=1/2$; all quantities are expressed in atomic units. 

\begin{figure}[htbp]
    \centering
    \begin{minipage}{0.48\textwidth}
        \centering
        \includegraphics[width=\linewidth]{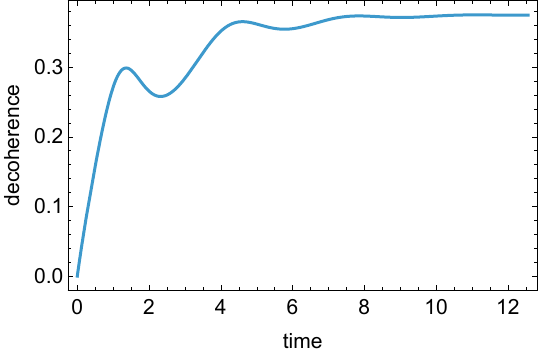}
        \centerline{(a)}
    \end{minipage}\hfill
    \begin{minipage}{0.48\textwidth}
        \centering
        \includegraphics[width=\linewidth]{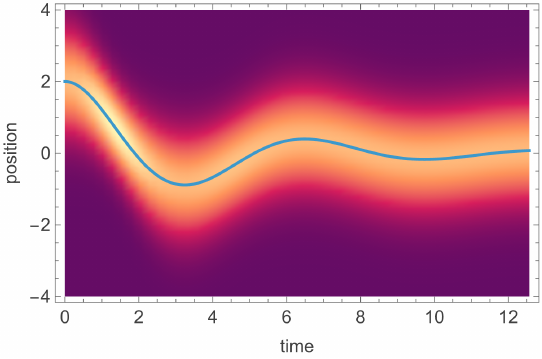}
        \centerline{(b)}
    \end{minipage}
    
    \caption{Exact propagation of a reduced density matrix simulating an open quantum system coupled to a harmonic environment. Panel (a) plots the time-dependent decoherence function, illustrating the continuous loss of quantum coherence as the system evolves. Panel (b) shows the exact spatio-temporal probability density $\rho(x,t)$ overlaid with the position center parameter $x_t$ (solid line). The analytical evaluation captures the exact spatial broadening and environmental damping dictated by the Caldeira-Leggett master equation.}
    \label{fig:rdm1d_dynamics}
\end{figure}

By querying the \texttt{"Decoherence"} property, users can explicitly track the continuous loss of quantum coherence as the system interacts with the thermal bath (full mathematical details are provided in the Supplementary Information). Simultaneously, extracting the spatial probability density (\texttt{"RHOX"}) and the position expectation value (\texttt{"RX"}) yields the exact relaxation of the system's spatial population and classical trajectory:  
\begin{lstlisting}[language=Mathematica, basicstyle=\footnotesize\ttfamily, columns=fullflexible, keepspaces=true, breaklines=true]
In[32]:= Needs["GWPTools`GWPEngineRDM1D`"]

In[33]:= OPT = "Potential" -> {"HarmonicCL", 1, 1/2, 1};
         OBJ9 = GWP["RDM1D"][1/4, 2, OPT];

In[34]:= pop[x_, t_] = OBJ9["RHOX"][x, t];
         dec[t_] = OBJ9["Decoherence"][t];
         rxt[t_] = OBJ9["RX"][t];
\end{lstlisting}
As illustrated in Figure \ref{fig:rdm1d_dynamics}, these analytical functions enable the exact visualization of the decoherence decay alongside the environmental damping and spatial broadening of the population. Beyond providing an intuitive pedagogical tool for demystifying open quantum systems, this exact algebraic representation lays the groundwork for advanced observable tracking. In future releases, this module will be expanded to evaluate quantum time-correlation functions, enabling the direct analytical prediction of spectroscopic signatures in dissipative environments.

\subsection{Multidimensional Dynamics and Mode Coupling}
\label{subsec:multidimensional}
Finally, the API established by the Parameter Bus is explicitly designed to scale toward higher-dimensional systems, akin to explicitly correlated Gaussian formulations \cite{Schrader2025}. Transitioning to two spatial dimensions requires a substantial expansion of the underlying wave packet dynamics. The prototype 2D engine (\texttt{GWPEngine2D.wl}) manages this by defining a 17-element Parameter Bus. The single 1D complex shape parameter expands into a $2\times2$ matrix tracking the real and imaginary components of the $xx$, $yy$, and cross-coupled $xy$ spatial spreads (\texttt{RAXX}, \texttt{IAXX}, \texttt{RAYY}, \texttt{IAYY}, \texttt{RAXY}, \texttt{IAXY}), while the position and momentum centers expand into explicit vectors (\texttt{RX}, \texttt{RY}, \texttt{RPX}, \texttt{RPY}). Completing this 17-element sequence are the global phase components (\texttt{RG}, \texttt{IG}), standard system constants (\texttt{MASS}, \texttt{HBAR}), and additional data (\texttt{NORM}, \texttt{PECOEFF}, \texttt{INIT}).

Because the architecture relies entirely on this generalized bus sequence rather than hardcoded dimensional rules, the frontend \texttt{GWPObject} container natively routes these 17 elements. However, to illustrate the raw power of the framework's modular backend, we bypass the frontend entirely and utilize the \texttt{GWPDeveloper} toolkit. We initialize a symmetric 2D wave packet (\texttt{PARAM0}) and directly propagate it through two distinct analytical models: an uncoupled 2D harmonic oscillator (\texttt{PARAM1}) and a coupled oscillator (\texttt{PARAM2}) with a cross-term potential coupling of $1/2$. For this demonstration, the wave packet is initialized at spatial coordinates $(x_0, y_0) = (-1, 0)$ with an initial momentum vector of $(p_{x0}, p_{y0}) = (0, -1)$:

\begin{lstlisting}[language=Mathematica, basicstyle=\footnotesize\ttfamily, columns=fullflexible, keepspaces=true, breaklines=true]
In[35]:= Needs["GWPTools`GWPEngine2D`"] 
         Get["GWPTools`GWPDeveloper`"]

In[36]:= PARAM0 = GWP2DPARAM[{{1/4, 0}, {0, 1/4}}, {-1, 0}, {0, -1}];
         PARAM1[t_] = GWP2DHARMONIC[1, 1, 0][t]@PARAM0;
         PARAM2[t_] = GWP2DHARMONIC[1, 1, 1/2][t]@PARAM0;
\end{lstlisting}
By extracting specific elements from the resulting parameter buses, we can simultaneously track the classical guiding center (\texttt{RXY}, elements 7 and 8) and the exact spatial probability covariance. The spatial covariance matrix is constructed algebraically by extracting the real width components (\texttt{RAXX}, \texttt{RAYY}, \texttt{RAXY} corresponding to elements 1, 3, and 5):

\begin{lstlisting}[language=Mathematica, basicstyle=\footnotesize\ttfamily, columns=fullflexible, keepspaces=true, breaklines=true]
In[37]:= {RAXX, RAYY, RAXY} = {PARAM2[t]}[[{1, 3, 5}]];
         RXY = {PARAM2[t]}[[{7, 8}]];
         COV = Inverse[2*{{RAXX, RAXY}, {RAXY, RAYY}}];
\end{lstlisting}

\begin{figure}[htbp]
    \centering
    \begin{minipage}{0.48\textwidth}
        \centering
        \includegraphics[width=\linewidth]{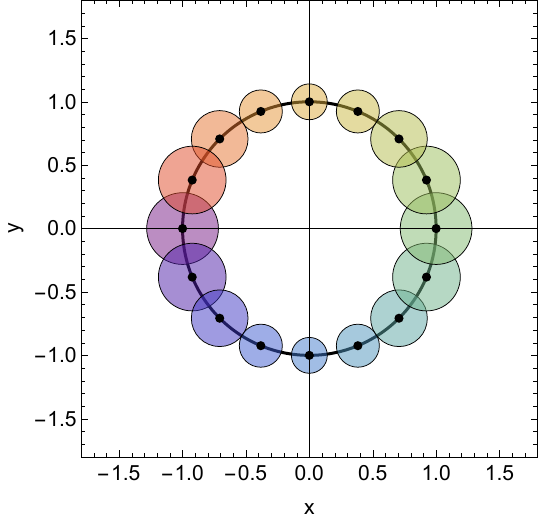}
        \centerline{(a)}
    \end{minipage}\hfill
    \begin{minipage}{0.48\textwidth}
        \centering
        \includegraphics[width=\linewidth]{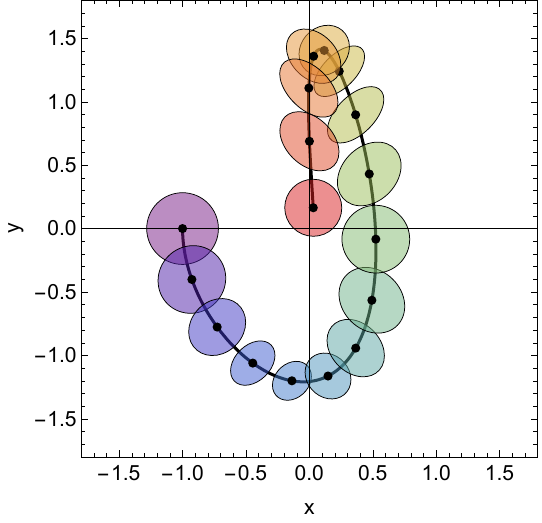}
        \centerline{(b)}
    \end{minipage}
    
\caption{Parametric trajectories of the 2D Gaussian wave packet overlaid with exact 1$\sigma$ spatial probability contours (ellipses) scaled by 1/25. Data are evaluated algebraically over a $2\pi$ temporal sequence via the experimental \texttt{GWPEngine2D} backend. Panel (a) shows the evolution within an uncoupled 2D harmonic oscillator. The wave packet maintains an isotropic profile as its guiding center traces a standard periodic circular orbit. Panel (b) shows the evolution within a coupled 2D harmonic oscillator. The spatial cross-coupling term in the potential induces continuous quantum squeezing and rotation of the probability density as it propagates.}
    \label{fig:2d_dynamics}
\end{figure}

While the positional elements trace out exact classical equations of motion, tracking the quantum covariance matrix reveals that the wave packet structurally twists and develops strong cross-spatial correlations as it evolves through the coupled potential. As visualized in Figure \ref{fig:2d_dynamics}, these algebraic components can be fed directly into standard geometric primitives to generate exact 1$\sigma$ probability contours. Pedagogically, this provides a highly transparent, exact model for visualizing vibrational energy transfer and mode coupling. Moving forward, the 2D architecture is uniquely positioned to incorporate explicitly time-dependent forcing fields and localized population and kinetic energy analysis. By systematically tuning the potential parameters and introducing external driving terms, this framework will serve as a rigorous analytical playground for exploring coherent control and driven multidimensional quantum dynamics.

\section{Conclusions}
\label{sec:conclusions}

In this work, we introduced \texttt{GWPTools}, a highly extensible \textit{Wolfram Language} framework for exact time-dependent Gaussian wave packet dynamics. Version 1.0.0 establishes a robust analytical core capable of evaluating exact algebraic properties and generating high-precision numerical trajectories for driven quadratic systems. By completely decoupling the frontend user interface from the backend mathematical evaluators via the \texttt{GWPDispatcher} and a structured Parameter Bus architecture, the software serves as an intuitive pedagogical tool, a reliable benchmarking engine for numerical methodologies, and a foundation for advanced theoretical prototyping.

As demonstrated by the advanced prototypes in Section \ref{sec:future_directions}, the architectural scaffolding natively supports advanced state formulations, dissipative environments, and multidimensional dynamics. By eliminating the accumulation errors inherent to discretized grids, \texttt{GWPTools} provides a zero-error reference standard. Ultimately, this unified symbolic-numeric infrastructure lays the foundation for sequenced non-linear spectroscopy calculations. Furthermore, a planned companion Python package will expand this framework's accessibility, paving the way for a comprehensive, scalable suite of exact quantum dynamics tools for researchers, students, and the broader open-source community.

\section*{Acknowledgements}
J.\ B.\ Maddox gratefully acknowledges travel support from Prof.\ Bill Poirier and the kind hospitality of his research group during a sabbatical at the University of Vermont (Summer and Fall 2025), where early versions of this package were developed. J.\ B.\ Maddox also wishes to thank Dr.\ Aric Hackebill (University of Vermont), Prof.\ Richard Lombardini (St.\ Mary's University), and Prof.\ Mikhail Khenner (Western Kentucky University) for many interesting conversations and encouraging feedback during the development of Version 1.0.0. Additionally, the authors extend their gratitude to Western Kentucky University for providing the institutional site license for Wolfram Mathematica, which was essential for this computational work.

\section*{Funding}
This research did not receive any specific grant from funding agencies in the public, commercial, or not-for-profit sectors.   

\section*{Declaration of Competing Interests}
The authors declare that they have no known competing financial interests or personal relationships that could have appeared to influence the work reported in this paper.   

\section*{Declaration of generative AI and AI-assisted technologies in the manuscript preparation process}
During the preparation of this work the author(s) used Google Gemini in order to refine manuscript phrasing, format LaTeX code, and assist in refining the software source code. After using this tool/service, the author(s) reviewed and edited the content as needed and take(s) full responsibility for the content of the published article.

\section*{Data Availability}
The \texttt{GWPTools} software package, including the source code, integrated documentation, and the automated \texttt{GWPVerification.nb} diagnostic suite, is openly available under the MIT License. The repository is hosted on GitHub at \url{https://github.com/maddoxjb/GWPTools}. The framework can be installed directly within a \textit{Wolfram Language} environment by evaluating the native package manager command with the direct release URL: \texttt{PacletInstall["https://github.com/maddoxjb/GWPTools/releases/download/v1.0.0/GWPTools-1.0.0.paclet"]}. Alternatively, developers can clone the repository and install the framework directly from the local directory structure.

\bibliographystyle{elsarticle-num} 
\bibliography{GWPTools_CPC}

\end{document}